\documentstyle[prl,aps,epsf,floats]{revtex}

\begin{document}
\draft

\twocolumn[\hsize\textwidth\columnwidth\hsize\csname @twocolumnfalse\endcsname
\title{Efficient nonlinear room-temperature spin injection from ferromagnets
into semiconductors through a modified Schottky barrier}  
\author{ V.V. Osipov and A.M. Bratkovsky }
\address{
Hewlett-Packard Laboratories, 1501 Page Mill Road, 1L, Palo Alto, CA 94304}
\date{ July 1, 2003 }
\maketitle
\begin{abstract}

We suggest a consistent microscopic theory of spin injection from a ferromagnet (FM) 
into a semiconductor (S). It describes tunneling and emission of 
electrons through modified FM-S Schottky barrier with an ultrathin heavily doped
interfacial S layer . We calculate nonlinear spin-selective  properties of such a 
reverse-biased FM-S junction, its nonlinear I-V characteristic, current saturation, 
and  spin accumulation in S. We show that the spin polarization of current, 
spin density, and penetration length increase with the total current 
until  saturation.  We find  conditions for most efficient spin injection, which 
are opposite to the results of previous works, since the present theory suggests
 using a lightly doped resistive semiconductor. It is shown that the maximal spin 
polarizations of current and electrons (spin accumulation) can approach 100\% 
at room temperatures and low current density in a nondegenerate high-resistance 
semiconductor.

\end{abstract}
\vskip2pc] \narrowtext
Spin injection into semiconductors holds promise for the next generation of
high-speed low-power electronic devices and perhaps even quantum computing 
\cite{Wolf,Datta,Hot,OB}. Among practically important spintronic effects is
a giant magnetoresistance in magnetic multilayers and tunnel
ferromagnet-insulator-ferromagnet (FM-I-FM) structures \cite{GMR,Slon,Brat}.
Injection of spin-polarized electrons into semiconductors is of particular
interest because of relatively large spin relaxation time \cite{Wolf}. An
efficient spin injection in heterostructures with magnetic semiconductor
(MS)\ as a spin source has been reported in Refs.~\cite{MSemi}. However, the
magnetization in MS\ usually vanishes or is too small at room temperature.
Relatively high spin injection from ferromagnets (FM) into nonmagnetic
semiconductors (S) has been recently demonstrated at low temperatures \cite
{Jonk}, the attempts to achieve an efficient room-temperature spin injection
have faced substantial difficulties \cite{Ferro}. Theoretical aspects of the
spin injection have been studied in Refs. \cite
{Aron,Mark,Son,Sch00,Flat,Alb,Her,Rash,Fert,Flat1,BOextr}.

Principal difficulty of the spin injection is that the materials in FM-S
junction\ usually have very different electron affinity and, therefore, a
high potential Schottky barrier forms at the interface \cite{sze}, Fig.~1.
For GaAs and Si the barrier height $\Delta \simeq 0.5-0.8$ eV with
practically all metals, including Fe, Ni, and Co, \cite{sze,Jonk} and the
barrier width is large, $l\gtrsim 30$ nm for doping concentration $%
N_{d}\lesssim 10^{17}$cm$^{-3}$. The spin-injection corresponds to a reverse
current in the Schottky contact, which is saturated and usually negligible
due to such large $l$\ and $\Delta $\ \cite{sze}. Therefore, a thin heavily
doped $n^{+}-$S layer between FM metal and S are used to increase the
reverse current \cite{sze} determining the spin-injection \cite{Jonk,Alb,OB}%
. This layer sharply reduces the thickness of the barrier, and increases its
tunneling transparency \cite{sze,OB}. Thus, a substantial spin injection has
been observed in FM-S junctions with a\ thin $n^{+}-$layer \cite{Jonk}.
However, the parameters of the structure \cite{Jonk} are not optimal (see
below).

A customarily overlooked paradox of spin injection is that a current through
Schottky junctions in prior theories depends solely on parameters of a
semiconductor \cite{sze} and cannot formally be spin-polarized. Some authors
even emphasize that in Schottky junctions ``spin-dependent effects do not
occur'' \cite{Sch00}.\ 

In earlier works \cite{Aron,Mark,Son,Sch00,Flat,Alb,Her,Rash,Fert,Flat1}
spin transport through FM-S junction, its spin-selective properties, and
nonlinear I-V characteristics have not been actually calculated. They were
described by various, often contradictory, boundary conditions at the FM-S
interface. For example, Aronov and Pikus assumed that a spin polarization of
current $P_{J}$ at the FM-S interface is a constant, equal to that in the
FM, and studied nonlinear spin accumulation in S considering spin diffusion
and drift in electric field \cite{Aron}. The authors of Refs. \cite
{Mark,Son,Sch00,Flat,Alb} assumed a continuity of both the currents and the
electrochemical potentials for both spins\ and found that a spin
polarization of injected electrons depends on a ratio of conductivities of a
FM and S (the so-called ``conductivity mismatch''\ problem). At the same
time, it has been asserted in Refs.~\cite{Her,Rash,Fert,Fert,Flat1} that the
spin injection becomes appreciable when the electrochemical potentials have
a substantial discontinuity (produced by e.g. a tunnel barrier \cite{Rash}).
The effect, however, was described by the unknown spin-selective interface
conductance $G_{i\sigma }$, which cannot be found within those theories.

We present a microscopic calculation of the spin transport through a
reverse-biased FM-S junction which includes an ultrathin heavily doped
semiconductor layer ($\delta -$doped layer) between FM and S. We find
conditions for the most efficient spin injection, which are opposite to the
results of \ previous phenomenological theories. We show that (i) the
current of the FM-S junction does depend on spin parameters of the
ferromagnetic metal but {\em not} its conductivity, so, contrary to the
results \cite{Mark,Son,Sch00,Flat,Alb,Rash,Fert,Flat1}, the ``conductivity
mismatch''\ problem {\em does not arise}. We find also that (ii) a spin
polarization of current $P_{J}$ of the FM-S junction strongly depends on the
current, contrary to the results \cite{Aron}, and (iii) the highest spin
polarization (close to 100\%) of both the injected electrons $P_{n}$ and
current $P_{J}$\ can be realized at room temperatures and relatively small
currents in {\em high-resistance} semiconductors, unlike claimed in Ref. 
\cite{Flat}, which are of most interest for spin injection devices \cite
{Datta,Hot,OB}. We show that (iv) tunneling resistance of the FM-S junction
has to be relatively small, which is {\em opposite }to the condition
obtained in linear approximation in Ref.\cite{Rash}. We find that (v) the
parameters $G_{i\sigma }$ are not constants, as was assumed in \cite
{Her,Rash,Fert,Flat1}, but vary with a current $J$ in a nonlinear fashion.

The necessary $\delta -$doped layer should be formed at the interface by
sequential donor and acceptor doping. The donor and acceptor concentrations, 
$N_{d}^{+}$ and $N_{a}^{+}$, and the corresponding thicknesses of the doping
profile, $l$ and $w$, have to satisfy the conditions: 
\begin{equation}
N_{d}^{+}l^{2}q^{2}\simeq 2\varepsilon \varepsilon _{0}(\Delta -\Delta
_{0}-rT),\text{ \ }N_{a}^{+}w^{2}q^{2}\simeq 2\varepsilon \varepsilon _{0}rT,
\label{Con0}
\end{equation}
and $l\lesssim l_{0},$ where $l_{0}=\sqrt{\hbar ^{2}/[2m_{\ast }(\Delta
-\Delta _{0})]}$ ($l_{0}\lesssim 2$ nm), $\Delta _{0}=(E_{c0}-F)>0$, $E_{c0}$
the bottom of conduction band in S in equilibrium, $q$ the elementary
charge, $\varepsilon $ ($\varepsilon _{0})$ the dielectric permittivity of S
(vacuum), $T$ the temperature in units of $k_{B}=1$ and $r\simeq 2-3$ (see
below). Thus, we consider a nondegenerated semiconductor and show that is
the case when the most efficient spin injection is realized at room
temperatures. A value of $\Delta _{0}$ can be set by choosing a donor
concentration in S, 
\begin{equation}
N_{d}=N_{c}\exp [(F-E_{c0})/T]=N_{c}\exp (-\Delta _{0}/T)=n,  \label{Nn}
\end{equation}
where $N_{c}=2M_{c}(2\pi m_{\ast }T)^{3/2}h^{-3}$ the effective density of
states and $M_{c}$ the number of effective minima of the semiconductor
conduction band; $n$ and $m_{\ast }$ the concentration and effective mass of
electrons in S \cite{sze}. The energy band diagram of such a FM-S junction
includes a potential $\delta -$spike of a height $(\Delta -\Delta _{0})$ and
a thickness $l,$ then a shallow potential well of a thickness $w\lesssim l$
and a depth of about $rT,$ followed by low and wide barrier with height $%
\Delta _{0}$ in the $n-$S, Fig.~1, i.e. $r$ is the ``mini-well'' parameter.\
Owing to small $l$, the electrons can rather easily tunnel through the $%
\delta -$spike but only those with an energy $E\geq E_{c}$ can overcome the
wide barrier $\Delta _{0}$ due to thermionic emission, where $%
E_{c}=E_{c0}-qV $ and $V>0$ is the bias voltage of the reversed-biased FM-S
junction. Presence of the mini-well allows to\ keep the thickness of the $%
\delta -$spike barrier equal to $l\lesssim l_{0}$ and its transparency high
at voltages $qV\lesssim rT$\ (see below). 
\begin{figure}[t]
\epsfxsize=3.4in 
\epsffile{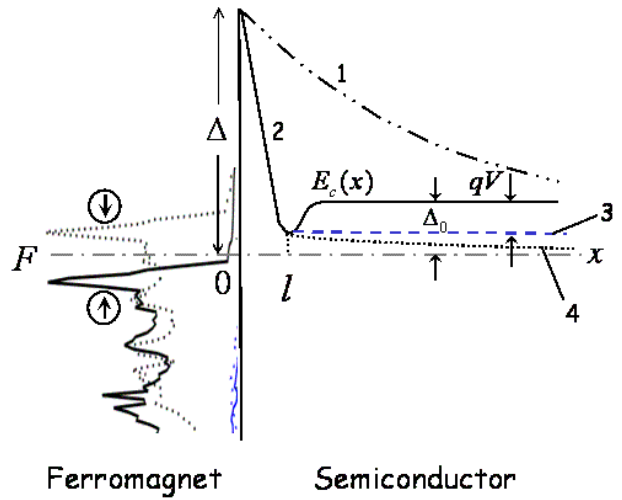}
\caption{Energy diagrams of ferromagnet-semiconductor heterostructure with $%
\protect\delta -$doped layer ($F$\ is the Fermi level; $\Delta $\ the height
and $l$\ the thickness of an interface potential barrier; $\Delta _{0}$\ the
height of the thermionic barrier in $n-$semiconductor). The standard
Schottky barrier (curve 1); $E_{c}(x)$\ the bottom of conduction band in $n-$%
semiconductor in equilibrium (curve 2), under small (curve 3), and large
(curve 4) bias voltage. The spin polarized density of states in Ni is shown
at $x<0$.}
\label{fig:fig1}
\end{figure}

We assume the elastic coherent tunneling, so that the energy $E$, spin $%
\sigma ,$ and $\vec{k}_{\parallel }$ (the component of the wave vector $\vec{%
k}$ parallel to the interface) are conserved. The current density of
electrons with spin $\sigma =\uparrow ,\downarrow $ through the FM-S
junction containing the $\delta -$doped layer (at the point $x=l$, Fig.~1)
can be written as \cite{Duke,Brat} 
\begin{equation}
J_{\sigma 0}=\frac{-q}{(2\pi )^{3}}\int d^{3}k[f(E_{k\sigma }-F_{\sigma
0}^{FM})-f(E_{k\sigma }-F_{\sigma 0})]v_{\sigma x}T_{k\sigma },  \label{amb1}
\end{equation}
where $T_{k\sigma }$ is the transmission probability, $f(E-F)$ the Fermi
function, $v_{\sigma x}$ the $x-$component of velocity\ $v_{\sigma }=\hbar
^{-1}|\nabla _{k}E_{k\sigma }|$ of electrons with spin $\sigma $\ in the
ferromagnet in a direction of current; the integration includes a summation
with respect to a band index. As distinct from Refs. \cite{Duke,Brat}, here
we study a strong {\em spin accumulation} in the semiconductor. Therefore,
we use {\em nonequilibrium} Fermi levels, $F_{\sigma 0}^{FM}$ and $F_{\sigma
0}$, describing distributions of electrons with spin $\sigma =\uparrow
,\downarrow $ in the FM and the S, respectively.{\bf \ }This approach is
valid when the spin relaxation time $\tau _{s}$ is much larger than the
relaxation time of electron energy, $\tau _{\epsilon },$ which is met in
practically all semiconductors at room temperature. In particular, the
electron density with spin $\sigma $\ in the S at the FM-S junction\ is
given by 
\begin{equation}
n_{\sigma 0}=(1/2)N_{c}\exp [(F_{\sigma 0}-E_{c})/T],  \label{nc}
\end{equation}
where $F_{\sigma 0}$ is a quasi Fermi level at a point $x=l$, Fig.~1. One
can see from (\ref{amb1}) that the current $J_{\sigma 0}=0$ if we take $%
F_{\sigma 0}^{FM}=F_{\sigma 0}$, i.e. use the supposition of Refs. \cite
{Mark,Son,Sch00,Flat,Alb}. In reality, due to very high electron density in
FM metal in comparison with electron density in S, $F_{\sigma 0}^{FM}$
extremely small differ from equilibrium Fermi level $F$ for currents under
consideration, therefore, as well as in Refs. \cite{Duke,Brat}, we can
assume that $F_{\sigma 0}^{FM}=F$ (see below a discussion). Then, we can be
rewritten Eq. (\ref{amb1}) as 
\begin{equation}
J_{\sigma 0}=\frac{q}{h}\int dE[f(E-F_{\sigma 0})-f(E-F)]\int \frac{%
d^{2}k_{\parallel }}{(2\pi )^{2}}T_{\sigma },  \label{amb}
\end{equation}
\qquad The current (\ref{amb}) should generally be evaluated numerically for
a complex band structure $E_{k\sigma }$\cite{stefano99}. The analytical
expressions for $T_{\sigma }(E,k_{\parallel })$\ can be obtained in an
effective mass approximation, $\hbar k_{\sigma }=m_{\sigma }v_{\sigma }$.
This applies to ``fast'' free-like d-electrons in elemental ferromagnets 
\cite{Stearns,Brat}. The present barrier has a ``pedestal'' with a height $%
(E_{c}-F)=\Delta _{0}-qV,$ which is opaque at energies $E<E_{c}$. For $%
E>E_{c}$ we approximate the $\delta -$barrier by a triangular shape and find 
\begin{equation}
T_{\sigma }={\frac{16\alpha m_{\sigma }m_{\ast }k_{\sigma x}k_{x}}{{m_{\ast
}^{2}k_{\sigma x}^{2}+m_{\sigma }^{2}}\kappa ^{2}}}e^{-\eta \kappa l}=\frac{%
16\alpha v_{\sigma x}v_{x}}{v_{\sigma x}^{2}+v_{tx}^{2}}e^{-\eta \kappa l},
\label{ts}
\end{equation}
where $\kappa =\left( 2m_{\ast }/\hbar ^{2}\right) ^{1/2}(\Delta
+F+E_{\parallel }-E)^{3/2}/(\Delta -\Delta _{0}+qV)$, $E_{\parallel }=\hbar
^{2}k_{\parallel }^{2}/2m_{\ast }$, $v_{x}=\sqrt{2(E-E_{c}-E_{\parallel
})/m_{\ast }}$ is the $x-$component of the velocity of electrons in S, $%
\hbar k_{x}=v_{x}m_{\ast }$, $v_{\sigma }=\hbar k_{\sigma }/m_{\sigma }$ the
velocity of electrons in the FM, $v_{t}=\hbar \kappa /m_{\ast }$ the
``tunneling'' velocity, $\alpha =\pi (\kappa l)^{1/3}\left[ 3^{1/3}\Gamma
^{2}\left( \frac{2}{3}\right) \right] ^{-1}\simeq 1.2(\kappa l)^{1/3},$ $%
\eta =4/3$ (for comparison, for a rectangular barrier $\alpha =1$ and $\eta
=2$).

The preexponential factor in Eq.~(\ref{ts}) takes into account a mismatch
between effective mass, $m_{\sigma }$ and $m_{\ast }$, and velocities, $%
v_{\sigma x}$ and $v_{x}$, of electrons in the FM and the S. We consider a
bias voltage $qV<\Delta _{0}$ and $T<\Delta _{0}\ll \Delta $. Taking also
into account that $v_{x}$ is nonzero only for the energy $E\geq
E_{c}=(F+\Delta _{0}-qV)>F,$ \ when electrons obey a classical statistics,
the considered S is nondegenerated. Therefore, using (\ref{ts}) the Eqs.~(%
\ref{amb}) can be written as 
\begin{eqnarray}
J_{\sigma 0} &=&{\frac{2^{5/2}\alpha qM_{c}}{{\pi hm_{\ast }^{1/2}}}}\left(
e^{F_{\sigma 0}/T}-e^{F/T}\right) \int_{0}^{\infty }dk_{\parallel
}^{2}\int_{E_{c}+E_{\parallel }}^{\infty }dE  \nonumber \\
&&\times {\frac{v_{\sigma x}\sqrt{E-E_{c}-E_{\parallel }}}{{v_{\sigma
x}^{2}+v_{tx}^{2}}}}\exp \left( -\eta \kappa l-\frac{E}{T}\right) .
\label{JSf}
\end{eqnarray}
Main contributions to the integral come from narrow intervals $%
0<E-E_{c}\lesssim T$ and $k_{\parallel }\lesssim \sqrt{m_{\ast }T}/\hbar $.
In this region the variables $v_{\sigma x}$, $v_{tx},$ and $\kappa $ are
smooth functions and can be replaced by constants with the result 
\begin{eqnarray}
J_{\sigma 0} &=&{\frac{2^{3/2}\alpha qM_{c}m_{\ast }^{1/2}v_{\sigma 0}T^{5/2}%
}{{\pi ^{3/2}\hbar ^{3}(v_{t0}^{2}+v_{\sigma 0}^{2})}}}\exp (-\eta \kappa
_{0}l)  \nonumber \\
&&\times \left( \exp {\frac{F_{\sigma 0}-E_{c}}{{T}}}-\exp \frac{F-E_{c}}{{T}%
}\right) ,  \label{J}
\end{eqnarray}
where $\kappa _{0}\equiv 1/l_{0}=(2m_{\ast }/\hbar ^{2})^{1/2}(\Delta
-\Delta _{0}+qV)^{1/2}$, $v_{t0}=\sqrt{2(\Delta -\Delta _{0}+qV)/m_{\ast }}$
and $v_{\sigma 0}=v_{\sigma }(E_{c})$. With the use of Eqs.~(\ref{Nn}) and (%
\ref{nc}) we obtain at $qV\lesssim rT$ 
\begin{eqnarray}
J_{\sigma 0} &=&j_{0}d_{\sigma }\left( \frac{2n_{\sigma 0}}{n}-\exp \frac{qV%
}{T}\right) ,  \label{JS} \\
\text{ }j_{0} &=&\alpha _{0}nqv_{T}\exp (-\eta \kappa _{0}l).  \label{J0}
\end{eqnarray}
We have introduced $\alpha _{0}=1.2(\kappa _{0}l)^{1/3}$, the thermal
velocity $v_{T}=\sqrt{3T/m_{\ast }}$ and the main spin factor 
\begin{equation}
d_{\sigma }=\frac{v_{T}v_{\sigma 0}}{v_{t0}^{2}+v_{\sigma 0}^{2}}  \label{ds}
\end{equation}
which determines the current dependence on materials constants of a
ferromagnet.

One can see from Eq. (\ref{JS}) that\ the total current $J=J_{\uparrow
0}+J_{\downarrow 0}$ and its spin components $J_{\sigma 0}$ depend on a\
conductivity of the semiconductor but {\em not} the ferromagnet, as in usual
Schottky junction theories \cite{sze}. However, $J_{\sigma 0}$ is
proportional to the spin factor $d_{\sigma }$ and the coefficient $%
j_{0}d_{\sigma }$ $\propto v_{T}^{2}\propto T$, but not the usual
Richardson's factor $T^{2}$ \cite{sze}.

Expression (\ref{JS}) is also valid at $V<0$ and determines the spin
current\ from S into FM. At $qV<-2T$ the spin current $J_{\sigma 0}\propto
n_{\sigma 0}$ and results in {\em extraction} of spin from nonmagnetic S
near the FM-S interface \cite{BOextr}.

At $qV\simeq rT\simeq (2-3)T$\ the value of $\exp (qV/T)\gg 2n_{\sigma 0}/n$
and, according to Eq.~(\ref{JS}), the spin polarization of the current, $%
P_{F}$, and the spin current at FM-S junction are equal, respectively, 
\begin{eqnarray}
P_{F} &=&\frac{J_{\uparrow 0}-J_{\downarrow 0}}{J_{\uparrow 0}+J_{\downarrow
0}}=\frac{d_{\uparrow }-d_{\downarrow }}{d_{\uparrow }+d_{\downarrow }} 
\nonumber \\
&=&\frac{(v_{\uparrow 0}-v_{\downarrow 0})(v_{t0}^{2}-v_{\uparrow
0}v_{\downarrow 0})}{(v_{\uparrow 0}+v_{\downarrow
0})(v_{t0}^{2}+v_{\uparrow 0}v_{\downarrow 0})},  \label{PF}
\end{eqnarray}
\begin{equation}
J_{\uparrow 0}=(1+P_{F})J/2.  \label{Jot}
\end{equation}
We remind that here $v_{\sigma 0}=v_{\sigma }(E_{c})$ where $%
E_{c}=(E_{c0}-eV)$. Thus, $P_{F}$ depends on bias voltage $V$ and differs
from that in usual tunneling MIM structures \cite{Brat}, since in the
present structure $P_{F}$ refers to the electron states in FM {\em above}
the Fermi level, at $E=E_{c}>F$, i.e. the high-energy equilibrium electrons,
which may be highly polarized (see below). Starting with Aronov and Pikus 
\cite{Aron}, usually one assumes a boundary condition $J_{\uparrow
0}=(1+P_{FM})J/2$, where 
\begin{equation}
P_{FM}=\frac{J_{\uparrow }-J_{\downarrow }}{J}=\frac{\sigma _{\uparrow
}-\sigma _{\downarrow }}{\sigma }
\end{equation}
is spin polarization of {\em current} in FM. However, there is a spin
accumulation in the S and near FM-S boundary in the semiconductor $n_{\sigma
0}=n/2+\delta n_{\sigma 0},$ where $\delta n_{\sigma 0}$ is a nonlinear
function of the current $J,$ $\delta n_{\sigma 0}\propto J$ at small current 
\cite{Aron} (see also below). Therefore, the greater is $J$, the high is $%
\delta n_{\sigma 0}$ and the smaller is the current $J_{\sigma 0}$ [see Eq.~(%
\ref{JS})]. In other words, a negative feedback is realized which decreases
the current polarization $P_{j}$\ and makes it also a nonlinear function of $%
J$, as we show below. We show that the spin polarizations of the current, $%
P_{jo}$, and the electrons, $P_{n0}=\left[ n_{\uparrow }(0)-n_{\downarrow
}(0)\right] /n$ in the semiconductor near FM-S junctions essentially differ
and both are low at small bias voltage $V$ (and current $J)$ but {\em %
increase} with the current up to $P_{F}$. Moreover, $P_{F}$ can essentially
differ from $P_{FM},$ and\ can approach 100\%.

Indeed, the current in a spin channel $\sigma $ is given by the expression 
\cite{Aron,Flat1} 
\begin{equation}
J_{\sigma }=q\mu n_{\sigma }E+qD\nabla n_{\sigma },  \label{eq:Js}
\end{equation}
where\ $E$ the electric field; $D$ and $\mu $ is the diffusion constant and
mobility of the electrons, in considered nondegenerated semiconductors $D$
and $\mu $ do not depend on the electron spin $\sigma $. From conditions 
\begin{equation}
J(x)=\sum_{\sigma }J_{\sigma }={\rm const}\text{ and }n(x)=\sum_{\sigma
}n_{\sigma }={\rm const}  \label{Const}
\end{equation}
we find 
\begin{equation}
E(x)=J/q\mu n={\rm const}\text{ and }\delta n_{\downarrow }(x)=-\delta
n_{\uparrow }(x).  \label{En}
\end{equation}
Note the injection of spin polarized electrons from FM into S corresponds to
a reverse current in the Schottky FM-S junction, i.e. in the considered case 
$J<0$ and $E<0,$ Fig.~1. The spatial distribution of density of electrons
with spin $\sigma $ in the semiconductor is determined by the continuity
equation \cite{Aron,Flat} 
\begin{equation}
\frac{dJ_{\sigma }}{dx}=\frac{q\delta n_{\sigma }}{\tau _{s}},  \label{CC}
\end{equation}
With the use of Eqs. (\ref{eq:Js}) and (\ref{En}), we obtain the equation
for $\delta n_{\uparrow }(x)=-\delta n_{\downarrow }(x)$. Its solution
satisfying a boundary condition $\delta n_{\uparrow }\rightarrow 0$ at $%
x\rightarrow \infty ,$ is \cite{Aron,Flat1} 
\begin{eqnarray}
\delta n_{\uparrow }(x) &=&C\frac{n}{2}\exp \left( -\frac{x}{L}\right) ,
\label{dn} \\
L &=&\frac{1}{2}\left( L_{E}+\sqrt{L_{E}^{2}+4L_{s}^{2}}\right)  \nonumber \\
&=&\frac{L_{s}}{2J_{S}}\left( \left| J\right| +\sqrt{J^{2}+4J_{S}^{2}}%
\right) ,  \label{L} \\
J_{S} &\equiv &qDn/L_{s},  \label{Jss}
\end{eqnarray}
where $L_{s}=\sqrt{D\tau _{s}}$ and $L_{E}=\mu |E|\tau _{s}=L_{s}\left|
J\right| /J_{S}$ are the spin-diffusion and the spin drift lengths,
respectively. Here $C$ defines the degree of spin polarization of
nonequilibrium electrons, i.e. a spin {\em accumulation} in the
semiconductor near the interface, 
\begin{equation}
C=\frac{n_{\uparrow }(0)-n_{\downarrow }(0)}{n}=P_{n}(0)\equiv P_{n0}.
\label{eq:Pn00}
\end{equation}
By substituting (\ref{dn}) into Eqs.~(\ref{eq:Js}) and (\ref{JS}), we find
that 
\begin{equation}
J_{\uparrow 0}=\frac{J}{2}\left( 1+P_{n0}\frac{L}{L_{E}}\right) =\frac{J}{2}%
\frac{\left( 1+P_{F}\right) (\gamma -P_{n0})}{\gamma -P_{n0}P_{F}},
\label{eq:Judt}
\end{equation}
where $\gamma =\exp (qV/T)-1$. From Eq.~(\ref{eq:Judt}), one obtains a
quadratic equation for $P_{n}(0)$ with a physical solution, which can be
written fairly accurately as 
\begin{equation}
P_{n0}=\frac{P_{F}\gamma L_{E}}{\gamma L+L_{E}}.  \label{c}
\end{equation}
By substituting (\ref{c}) into Eqs.~(\ref{JS}) we find that the total
current $J=J_{\uparrow 0}+J_{\downarrow 0}$ can be written fairly accurately
as 
\begin{eqnarray}
J &=&-J_{m}\gamma =-J_{m}\left( e^{qV/T}-1\right) ,  \label{CT} \\
J_{m} &=&\alpha _{0}nqv_{T}(1-P_{F}^{2})(d_{\uparrow 0}+d_{\downarrow
0})e^{-\eta \kappa _{0}l}.  \label{Jm}
\end{eqnarray}

We notice that at a voltage $qV\simeq rT$ \ the shallow potential mini-well
vanishes and $E_{c}(x)$ takes the shape shown in Fig.~1 (curve 3). For $%
qV>rT $, a wide potential barrier at $x>l$ remains flat (characteristic
length scale $\gtrsim 100$nm at $N_{d}\lesssim 10^{15}$cm$^{-3}$)$,$ as in
usual Schottky contacts \cite{sze}. Therefore the current becomes weakly
depending on $V$, since the barrier is opaque for electrons with energies $%
E<E_{c}-rT$ (Fig.~1, curve 4). Thus, Eq. (\ref{CT}) is valid only at $%
qV\lesssim rT$ and the current at $qV\gtrsim rT$ practically saturates at
the value 
\begin{equation}
J_{sat}=qn\alpha _{0}v_{T}(d_{\uparrow 0}+d_{\downarrow 0})(1-P_{F}^{2})\exp
\left( r-\eta \kappa _{0}l\right) .  \label{Sat}
\end{equation}
With the using (\ref{CT}) and (\ref{L}), we obtain from Eq. (\ref{c}) that
spin polarization of electrons at FM-S juction is equal to 
\begin{equation}
P_{n0}=\frac{\left| J\right| L_{E}P_{F}}{J_{m}L_{E}+\left| J\right| L}=\frac{%
2\left| J\right| P_{F}}{2J_{m}+\left| J\right| +\sqrt{J^{2}+4J_{S}^{2}}}
\label{Pn0}
\end{equation}

The spin polarization of the current at FM-S interface, according to (\ref
{JS}), (\ref{eq:Judt}), (\ref{L}) and (\ref{Pn0}), is 
\begin{eqnarray}
P_{J0} &=&\frac{J_{\uparrow 0}-J_{\downarrow 0}}{J_{\uparrow
0}+J_{\downarrow 0}}=P_{n}(0)\frac{L}{L_{E}}  \nonumber \\
&=&P_{F}\frac{\left| J\right| +\sqrt{J^{2}+4J_{S}^{2}}}{2J_{m}+\left|
J\right| +\sqrt{J^{2}+4J_{S}^{2}}}.  \label{PJ0}
\end{eqnarray}

One can see that $P_{J0}$ strongly differs from $P_{n0}$ at small currents.
As expected $P_{n}\approx P_{F}\left| J\right| /J_{m}\rightarrow 0$ vanishes
with current, the constant of proportionality differs from those obtained in
Refs. \cite{Aron,Flat,Her,Fert,Flat1}. At large currents $J\gg J_{m}$ the
spin polarization (of electron density) approaches maximum value $P_{F}.$
Unlike the spin accumulation $P_{n0}$, the current polarization $P_{J0}$
does not vanish at small currents, but approaches a value $%
P_{J0}^{0}=P_{F}J_{S}/(J_{S}+J_{m})\ll P_{F}$ (see below). Both $P_{n0}$ and 
$P_{J0}$ approach the maximum $P_{F}$ only when $\left| J\right| \gg
J_{m},J_{S}$, Fig.~2. The condition $\left| J\right| \gg J_{m},J_{S}$ is
fulfilled at $qV\simeq rT\gtrsim 2T,$ when $J_{m}\gtrsim J_{S}.$ According
to (\ref{Jm}) and (\ref{Jss}), the condition $J_{m}\gtrsim J_{S}$ for
maximal polarization $P_{J}$ and $P_{n}$, can be written as 
\begin{equation}
\beta =\alpha _{0}v_{T}(d_{\uparrow 0}+d_{\downarrow
0})(1-P_{F}^{2})e^{-\eta l/l_{0}}\tau _{s}/L_{s}\gtrsim 1.  \label{Con1}
\end{equation}
It can be met most easily for a thin tunneling $\delta -$doped layer with $%
l\lesssim l_{0}$, and semiconductors with a long enough spin-relaxation
time. The condition (\ref{Con1}) is satisfied for typical parameters at $%
T\simeq 300$ K ($D\approx 25$ cm$^{2}$/s, $\Delta \simeq 0.5$ eV, and $%
v_{\sigma 0}\simeq 10^{8}$ cm/s) at $l\lesssim l_{0}\simeq 2{\rm nm},$ when $%
\tau _{s}\gg 10^{-13}$s. The spin-coherence time $\tau _{s}$ in conventional
semiconductors at room temperature is greater than $10^{-11}$s \cite{Wolf}
and can be as long as $\sim 1$ ns (e.g. in ZnSe \cite{kik1ns}). We notice,
that when $l\lesssim l_{0},$ the current spin polarization at small current $%
P_{J0}^{0}=P_{F}/(1+\beta )\ll P_{F}$, since in this case the value $\beta
\simeq (d_{\uparrow 0}+d_{\downarrow 0})\alpha _{0}v_{T}\tau _{s}/L_{s}\gg 1$
for real semiconductor parameters.

Note that the higher the semiconductor conductivity, $\sigma _{S}=q\mu
n\propto n$, the larger the threshold current $J>J_{m}\propto n$ for
achieving the maximal spin injection. In other words, the polarizations $%
P_{Jo}$ and $P_{no}$ reach the maximum $P_{F}$ at ${\em smaller}$ current
in\ {\em high-resistance} low-doped semiconductors rather than in heavily
doped semiconductors. Therefore, there is no such thing as a ``conductivity
mismatch'' \cite{Sch00,Rash,Fert}.

\begin{figure}[t]
\epsfxsize=3.4in 
\epsffile{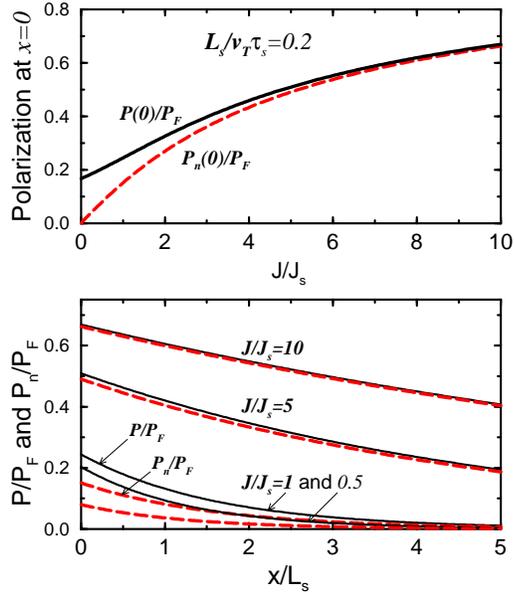}
\caption{ Spin polarization of a current $P=(J_{\uparrow }-J_{\downarrow
})/J $ (solid line) and spin accumulation $P_{n}=(n_{\uparrow
}-n_{\downarrow })/n $ (broken line) in the semiconductor as the functions
of the relative current density $J/J_{s}$ (top panel) and their spatial
distribution for different densities of total current $J/J_{s}$\ (bottom
panel).at $L_{s}/v_{T}\protect\tau _{s}=0.2,$ where $J_{s}=qnL_{s}/\protect%
\tau _{s}$, $P_{F}$ is the spin polarization in the ferromagnet (see text).}
\label{fig:fig2}
\end{figure}
According to Eq.~(\ref{CT}), at\ small voltages, $qV<T$ the necessary
condition $\left| J\right| \gg J_{s},$ can be rewritten as 
\begin{equation}
r_{c}\ll L_{s}/\sigma _{S}  \label{rc}
\end{equation}
where $r_{c}=(dJ/dV)^{-1}$is the tunneling contact resistance. Here we have
used the Einstein relation $D/\mu =T/q$ for nondegenerate semiconductors. We
emphasize that Eq.~(\ref{rc}) is {\em opposite} to the condition found in
Ref. \cite{Rash} for small currents. Indeed, at small currents $%
P_{J0}=P_{F}/(1+\beta )\simeq P_{F}$ only when\ $\beta \ll 1$, i.e. $%
r_{c}\gg L_{s}/\sigma _{S}$ (cf. Ref. \cite{Rash}). However, at such large
tunneling contact resistance $r_{c}$ the saturation current $J_{sat}$ of the
FM-S junction is much smaller than $J_{s}$. Therefore, the degree of spin
accumulation in the semiconductor is very small, $P_{n0}\ll 1$, but this $%
P_{n0}$ is exactly the characteristic that determines the main spin effects 
\cite{Wolf,Datta,Hot,OB}. Note that the conclusion (\ref{rc}) does not
depend on the electron concentration in the semiconductor, $n$, and is also
valid for heavily doped degenerate semiconductors.

We notice that the quasi Fermi level $F_{\sigma 0}^{FM}$ for electrons with
spin $\sigma $ in FM extremely small differ from equilibrium Fermi level $F$%
. It is easy to verify that $\left| F_{\sigma 0}^{FM}-F\right| \ll \left|
F_{\sigma 0}-F\right| $ at current $J\lesssim J_{sat}$ by virtue of
smallness of relation $n/n_{FM}\ll 1$, where $n_{FM}$ is electron density in
FM metal. Thus, above used assumption $F_{\sigma 0}^{FM}=F$ is valid.

According to Eq. (\ref{ds}) in the effective mass approximation the spin
factor $d_{\sigma }\propto v_{\sigma 0}^{-1}$, since usually $v_{\sigma
0}>v_{t0}$. In a metal $v_{\sigma 0}^{-1}\propto g_{\sigma 0}=g_{\sigma
}(E_{c}),$ so that $d_{\sigma }\propto g_{\sigma }(E_{c})$ where $g_{\sigma
0}=g_{\sigma }(E_{c})$ is the density of states of the d-electrons with spin 
$\sigma $ and energy $E=E_{c}$ in the ferromagnet. Thus, supposing $%
m_{\sigma }=m$ we find from Eq. (\ref{PF}) $P_{F}\approx (g_{\uparrow
0}-g_{\downarrow 0})/(g_{\uparrow 0}+g_{\downarrow 0})$. One assumes that
the same proportionality between the polarization and the density of states
approximately holds in general case of more complex band structures. Note
that the polarization of d-electrons in elemental ferromagnets Ni{\it ,} Co%
{\it \ }and\ Fe is reduced by the current of unpolarized s-electrons $rJ_{s}$%
, where $r<1$ is a factor (roughly the ratio of the number of s-bands to the
number of d-bands crossing the Fermi level). Together with the contribution
of s-electrons the total polarization is approximately

\begin{equation}
P_{F}=\frac{J_{\uparrow 0}-J_{\downarrow 0}}{J_{\uparrow 0}+J_{\downarrow
0}+J_{s0}}\simeq \frac{g_{\uparrow 0}-g_{\downarrow 0}}{g_{\uparrow
0}+g_{\downarrow 0}+2rg_{s0}},  \label{eq:Pol}
\end{equation}
Such a relation for $P_{F}$ can be obtained from a usual ``golden-rule''
type approximation for tunneling current (cf. Refs. \cite{sze,other,Esaki}).
The density of states $g_{\downarrow }$ for minority d-electrons in Fe, Co,
and Ni has\ a larger pick at $E=E_{F}+\Delta _{\downarrow }$ ($\Delta
_{\downarrow }\simeq 0.1$ eV) which is much larger than $g_{\uparrow }$ for
the majority $d-$electrons and $g_{s}$ for $s-$electrons \cite{Mor} (see
Fig. 1). The FM-S junction in Fig. 1 can be tailored to adjust the cutoff
energy $E_{c}\simeq E_{F}+\Delta _{\downarrow }$ to the peak in minority
electrons. Thus, if one selects $\Delta _{0}=\Delta _{\downarrow }+eV\simeq
\Delta _{\downarrow }+rT,$ then $g_{\downarrow 0}\gg g_{\uparrow 0}\gg $ $%
g_{s0},$ and, according to Eq. (\ref{eq:Pol}),\ the polarization\ $P_{F}$\
may be close to 100\%. (note, that in present case the polarization $P_{F}$
is negative, $P_{F}\approx -1$). \ 

We emphasize that the spin injection in structures \cite
{Hot,Jonk,Ferro,Aron,Mark,Son,Sch00,Flat,Alb,Her,Rash,Fert,Flat1} has been
dominated by electrons at the Fermi level and, according to calculation \cite
{Mor}, $g_{\downarrow }(F)$ and $g_{\uparrow }(F)$ are such that $%
P_{F}\lesssim 40\%$. \ We also notice that the condition (\ref{Con1}) for
parameters of the Fe/AlGaAs heterosructure studed in Refs. \cite{Jonk} ($%
l_{+}\simeq 3$ nm, $l_{0}\simeq 1$ nm and $\Delta _{0}=0.46$ eV) is
satistied when $\tau _{s}\gtrsim 5\times 10^{-10}$s, that can be fulfilled
only at low temperature. Moreover, for the concentration $n=10^{19}$cm$^{-3}$
$E_{c}$ lies below $F$ where $P_{F}\lesssim 40\%$. Therefore, the authors of
Refs. \cite{Jonk} were indeed able to observe spin polarization of electrons 
$P_{n}=32\%$ at low temperatures.

More controlled parameters can be realized in heterostructures in which $%
\delta -$ layer between the ferromagnet and the $n-$semiconductor layer made
from very thin highly doped $n^{+}$-semiconductor with larger electron
affinity than the $n-$ semiconductor. Such heterostructures can be ${\rm FM}%
-n^{+}$-${\rm GaAs}-n$-${\rm Ga}_{1-x}{\rm Al}_{x}{\rm As}$, ${\rm FM}-n^{+}$%
-${\rm Ge}_{x}%
\mathop{\rm Si}%
_{1-x}-n$-$%
\mathop{\rm Si}%
$ or ${\rm FM}-n^{+}$-${\rm Zn}_{1-x}{\rm Cd}_{x}{\rm Se}-n$-${\rm ZnSe}$
structures. The ${\rm GaAs}$, ${\rm Ge}_{x}%
\mathop{\rm Si}%
_{1-x}$ or ${\rm Zn}_{1-x}{\rm Cd}_{x}{\rm Se}$ $n^{+}-$layer must have the
width $l_{+}<1$ nm and the concentration $N_{d}^{+}>10^{20}$cm$^{-3}$. In
this case the superthin barrier forming near the ferromagnet-semiconductor
interface is transparent for the electron tunnelling. The barrier height $%
\Delta _{0}$ at ${\rm Ge}_{x}%
\mathop{\rm Si}%
_{1-x}-%
\mathop{\rm Si}%
$, ${\rm GaAs}-{\rm Ga}_{1-x}{\rm Al}_{x}{\rm As}$ or ${\rm Zn}_{1-x}{\rm Cd}%
_{x}{\rm Se}-{\rm ZnSe}$ interface is determined by the composition $x$ and
can be selected as $\Delta _{0}=0.05-0.15$eV. When the donor concentration
in $%
\mathop{\rm Si}%
$, ${\rm Ga}_{1-x}{\rm Al}_{x}{\rm As}$ or ${\rm ZnSe}$ layer $N_{d}<10^{17}$
cm$^{-3}$ the injected electron can not propagate through the low barrier $%
\Delta _{0}$ when its width $l_{0}>10$ nm.

Summarizing, we showed that (i) the most efficient spin injection at
room-temperature occurs in ferromagnetic-semiconductor junctions when an
ultrathin heavily $n^{+}-$doped semiconductor layer ($\delta -$doped layer)
satisfying certain conditions is formed between the ferromagnet and the
nondegenerate $n-$type semiconductor, (ii) the conduction band bottom of the
semiconductor, $E_{c}=E_{c0}+eV\simeq E_{c0}+rT$ should be close to the peak
in the density of minority electron states of elemental ferromagnets Ni{\it ,%
} Co{\it \ }and\ Fe; (iii) the reverse current of such modified Schottky
junctions, which determines the spin-injection from ferromagnets into
semiconductors, is determined by tunneling and thermoionic emission of spin
polarized electrons; (iv) spin injection depends on parameters of both a
semiconductor and a ferromagnet, in particular, on velocity of electrons
with spin $\sigma $ and energy $E\simeq E_{c}$, and a\ conductivity of the
semiconductor (but{\bf \ }{\em not}{\bf \ }the ferromagnet); (v) spin
polarizations of current, $P_{J0}$, and electrons, $P_{n0}$, in the
semiconductor differ from each another and are small at low current; they
increase with the total current and reach the maximal possible value\ $%
\left| P_{F}\right| \approx 1$ only at relatively large current $J_{m},$
when the spin penetration depth $L_{E}$ is much larger than the spin
diffusion length $L_{s}$, and (vi) the smaller the semiconductor
conductivity, the lower threshold current $J_{m}$ for achieving an efficient
spin injection. The present theory opens up the way of optimizing the
spin-injection structures towards achieving 100\% spin polarization in a
semiconductor at room temperatures.

We thank D.D. Awschalom, E.I.Rashba, and I. Zutic for useful discussions.


\end{document}